\documentclass[nolayout]{article}
\usepackage[english]{babel}
\usepackage{authblk}
\usepackage{graphicx,xcolor,xargs,twoopt,dsfont}
\usepackage{framed}
\usepackage[normalem]{ulem}
\usepackage[active]{srcltx}
\usepackage{amsmath}
\usepackage{amsthm}
\usepackage{amssymb}
\usepackage{amsfonts}
\usepackage{enumerate}
\usepackage[utf8]{inputenc}
\usepackage{fancyhdr}
\usepackage{geometry}
\usepackage{booktabs}
\newcommand{\eqsp}{\,}

\newcommandx\filtderiv[2][1=]{
\ifthenelse{\equal{#1}{}}
	{\eta_{#2}}
	{\eta_{#2}^\N}
}

\newcommand{\tstatletter}{\kernel{T}}

\newcommandx\tstat[2][1=]{
\ifthenelse{\equal{#1}{}}
	{\tstatletter_{#2}}
	{\tau_{#2}^{#1}}
}
\newcommandx\tstathat[2][1=]{
\ifthenelse{\equal{#1}{}}
	{\tstatletter_{#2}}
	{\widehat{\tau}_{#2}^{#1}}
}

\newcommand{\kernel}[1]{\mathbf{#1}}

\newcommandx{\bk}[2][1=]{ 
\ifthenelse{\equal{#1}{}}
{\overleftarrow{\kernel{Q}}_{#2}}
{\overleftarrow{\kernel{Q}}_{#2}^{#1}}
}

\newcommandx{\bkhat}[2][1=]{ 
\ifthenelse{\equal{#1}{}}
{\widehat{\kernel{Q}}_{#2}}
{\widehat{\kernel{Q}}_{#2}^{#1}}
}

\newcommand{\N}{N}

\newcommandx{\K}[1][1=]{
\ifthenelse{\equal{#1}{}}{{\kletter}}{{\widetilde{\N}^{#1}}}}

\newcommand{\kletter}{\widetilde{\N}}

\def\1{\mathds{1}}

\newcommand{\esssup}[2][]
{\ifthenelse{\equal{#1}{}}{\left\| #2 \right\|_\infty}{\left\| #2 \right\|^2_{\infty}}}

\newcommand{\kiss}[3][]
{\ifthenelse{\equal{#1}{}}{r_{#2|#3}}
{\ifthenelse{\equal{#1}{fully}}{r^{\star}_{#2|#3}}
{\ifthenelse{\equal{#1}{smooth}}{\tilde{r}_{#2|#3}}{\mathrm{erreur}}}}}

\newcommand{\chunk}[4][]%
{\ifthenelse{\equal{#1}{}}{\ensuremath{{#2}_{#3:#4}}}{\ensuremath{#2^#1}_{#3:#4}}
}

\newcommand{\kissforward}[3][]
{\ifthenelse{\equal{#1}{}}{p_{#2}}
{\ifthenelse{\equal{#1}{fully}}{p^{\star}_{#2}}
{\ifthenelse{\equal{#1}{smooth}}{\tilde{r}_{#2}}{\mathrm{erreur}}}}}

\newcommandx\post[2][1=]{
\ifthenelse{\equal{#1}{}}
	{\phi_{#2}}
	{\phi_{#2}^\N}
}

\newcommandx\posthat[2][1=]{
\ifthenelse{\equal{#1}{}}
	{\widehat{\phi}_{#2}}
	{\widehat{\phi}_{#2}^\N}
}

\newcommand{\adjfunc}[4][]
{\ifthenelse{\equal{#1}{}}{\ifthenelse{\equal{#4}{}}{\vartheta_{#2|#3}}{\vartheta_{#2|#3}(#4)}}
{\ifthenelse{\equal{#1}{smooth}}{\ifthenelse{\equal{#4}{}}{\tilde{\vartheta}_{#2|#3}}{\tilde{\vartheta}_{#2|#3}(#4)}}
{\ifthenelse{\equal{#1}{fully}}{\ifthenelse{\equal{#4}{}}{\vartheta^\star_{#2|#3}}{\vartheta^\star_{#2|#3}(#4)}}{\mathrm{erreur}}}}}

\newcommand{\XinitIS}[2][]
{\ifthenelse{\equal{#1}{}}{\ensuremath{\rho_{#2}}}{\ensuremath{\check{\rho}_{#2}}}}

\newcommand{\filt}[2][]%
{%
\ifthenelse{\equal{#1}{}}{\ensuremath{\phi_{#2}}}{\ensuremath{\phi_{#1,#2}}}%
}

\newcommand{\sumwght}[2][]{%
\ifthenelse{\equal{#1}{}}{\ensuremath{\Omega_{#2}}}{\ensuremath{\Omega_{#2}^{(#1)}}}}
\newcommand{\sumwghthat}[2][]{%
\ifthenelse{\equal{#1}{}}{\ensuremath{\widehat{\Omega}_{#2}}}{\ensuremath{\widehat{\Omega}_{#2}^{(#1)}}}}

\newcounter{hypH}

\begin{document}

\title{End-to-end deep metamodeling to calibrate and optimize energy loads}
\date{} 

\author[$\dag$]{Max Cohen}
\author[$\ddag$]{Maurice Charbit}
\author[$\dag$]{Sylvain Le Corff}
\author[$\star$]{Marius Preda}
\author[$\ddag$]{Gilles Nozi\`ere}
\affil[$\dag$]{{\small  Samovar, T\'el\'ecom SudParis, D\'epartement CITI, TIPIC, Institut Polytechnique de Paris.}}
\affil[$\star$]{{\small  Samovar, T\'el\'ecom SudParis, D\'epartement ARTEMIS, ARMEDIA, Institut Polytechnique de Paris.}}
\affil[$\ddag$]{{\small  Oze-\'Energies.}}

\lhead{Max Cohen et al.}
\rhead{End-to-end deep metamodeling to calibrate and optimize energy loads}

\maketitle

\begin{abstract}
In this paper, we propose a new end-to-end methodology to optimize the energy performance and the comfort, air quality and hygiene of large buildings. A metamodel based on a Transformer network is introduced and trained using a dataset sampled with a simulation program. Then, a few physical parameters and the building management system settings of this metamodel are calibrated using the CMA-ES optimization algorithm and real data obtained from sensors. Finally, the optimal settings to minimize the energy loads while maintaining a target thermal comfort and air quality are obtained using a multi-objective optimization procedure. The numerical experiments illustrate how this metamodel ensures a significant gain in energy efficiency while being computationally much more appealing than models requiring a huge number of physical parameters to be estimated.
\end{abstract}

\section{Introduction}
Global energy demand for heating, ventilation and air-conditioning  in commercial or public buildings has been increasing rapidly for the past few decades along with population and economic growth. This rising demand is at the root of the complex problem of simultaneously ensuring a better environmental impact, as higher consumption of fossil fuels implies higher greenhouse gas emissions, while maintaining a satisfactory comfort in buildings (air and indoor temperature quality). In that respect, building designing and management have to integrate thermal performance and comfort criteria, and to assess the environmental consequences of any chosen policy. This makes the analysis of building energy performance a challenging multi-criteria problem, as detailed for instance in \cite{Bre2016ResidentialBD}. Our paper sets the focus on analyzing and optimizing cooling, heating and air conditioning loads by tuning the building management system in given buildings without costly, invasive or time consuming renovation works. The aim is to provide the optimal building management settings, governing Heating, Ventilation and Air-Conditioning (HVAC) and Air Handling Units (AHU), in order to improve thermal comfort, energy loads as well as environmental impact. This objective is decomposed into three steps: (i) provide a model to predict future energy loads and temperature in a building based on the HVAC system and the weather forecast, (ii) calibrate the parameters of this model based on real data obtained in real time in each building and (iii) optimize the HVAC equipment to minimize the total energy load in future periods while maintaining a given thermal comfort.

The first category of approaches to model the energy performance of a building are based on physical equations that describe heat transfer between the building and its environment. Thanks to their increasing reliability, simulation based methods such as EnergyPlus, TRNSYS or DOE-2  are commonly used to simulate the system behavior based on a schematic view of the building. EnergyPlus was used for instance in \cite{shabunko2018energyplus} to build three types of typical designs and to benchmark the energy performance of 400 residential buildings. In \cite{zhao2016occupant}, the authors proposed a predictive control framework based on Matlab and EnergyPlus in order to optimize energy consumptions while meeting the individual thermal comfort preference. In these papers, a schematic building is used in the simulation program and considered as a baseline for energy loads. 
These approaches rely on a huge number of parameters, such as window to wall ratio, window leakage, or wall construction. Instead of costly campaigns to measure these parameters, that would have to be reiterated for each new building, they may be estimated using an automatic calibration procedure by minimizing a cost function which associates, with each set of parameters, the discrepancy between the true energy loads and temperatures, and the simulated ones, see \cite{Coakley2014ARO, Corff2018OPTIMIZINGTC}. As shown in \cite{Nagpal2019AMF}, calibration yields sufficiently accurate results for a variety of different buildings, thus ensuring limited additional costs to generalize a given model. Once calibrated, the optimisation task consists in determining a set of building management settings that will result in lower energy consumption, while preserving comfort. Following numerous works such as \cite{Bre2020AnEM}, the multi-objective Non-dominated Sorting Genetic Algorithm-II (NSGA-II), see \cite{Deb2000AFE}, is the most widespread method to solve the optimization task.  However, when no prior knowledge is available on the thousands of specific parameters required to specify each building, calibrating and optimizing such simulation programs is computationally prohibitive, see \cite{Westermann2019SurrogateMF}. This shortcoming is particularly severe in cases where many data are available from numerous wireless sensors installed in a building but no intrusive and resource consuming in-site campaigns are deployed to fix the values of the physical parameters.

Metamodeling approaches aim at overcoming this computational cost by proposing surrogate models that replace the physical simulator during calibration and optimization tasks. The parameters of such metamodels are estimated during a training phase using simulations conducted by a physical-based model, that aims at exhaustively capturing the building behavior for various building management settings.
In \cite{Bre2020AnEM, Reynolds2018AZB}, statistical models are trained on a dataset sampled from EnergyPlus, allowing significant computational savings during optimization. In \cite{Bre2020AnEM}, the authors proposed to combine NSGA-II with an artificial neural network metamodel to obtain a Pareto front of optimal HVAC parameters with the trained metamodel, in order to optimize the consumption of a $83\,\mathrm{m}^2$ house.  To fit this dataset, instead of standard statistical models, this paper uses a Feed Forward Neural Network (FFN) as a metamodel. Although they often yield very accurate predictions, these neural networks are not adapted to time series problems, and are usually substituted for there sequential counter parts, such as recurrent or convolutional based approaches. This FFN was only validated with EnergyPlus simulations, and the calibration step was not performed, as no real historic data from the targeted building were discussed. Similarly, \cite{Reynolds2018AZB} proposed a FFN based metamodeling approach to reduce up to 25\% the energy consumption in a small office building. EnergyPlus was used to sample a dataset for various zones of the building, in order to train zone level metamodels. These simulation spread over 24 hours, and approximated the building behavior from January to March. Once again, in the absence of real historic data, no calibration step was implemented. The first optimization method is similar to previous works, and consists in optimizing consumption by feeding NSGA-II each metamodel. In a novel approach, optimization can also be updated every hour with the newly collected data from the building, in hope to avoid the error drift of simulating 24 hours of building behavior without any feedback. The study presented in \cite{Magnier2010MultiobjectiveOO} focused on the optimization of a $210\,\mathrm{m}^2$ two storey house. Despite measured data being available, no automatic calibration step was discussed ; instead, TRNSYS was calibrated by hand to match the real building, resulting in relative errors of 3.7\%, 3.4\%, and 7.3\% for heating, cooling, and fan monthly energy consumption, respectively. A FFN surrogate model was fit on a dataset of 450 samples, before being optimized using once again NSGA-II. 

In this paper, we propose a end-to-end methodology, from dataset sampling to metamodel calibration and optimization using data obtained from wireless sensors set in a large building. The proposed metamodels involve classical recurrent neural networks and an approach based on a Transformer architecture (\cite{Vaswani2017AttentionIA}) which has recently proven both an accurate and computing efficient alternative to traditional sequences to sequences models, such as Long Short-Term Memory (LSTM, \cite{Hochreiter1997LongSM}) and Gated Recurrent Unit (GRU, \cite{Cho2014LearningPR}). Transformers combine an encoder-decoder architecture, see for instance \cite{Cho2014LearningPR} or \cite{Bahdanau2014NeuralMT}, allowing the model to learn semantic information from the observations using attention mechanisms (\cite{Parikh2016ADA, Zhu2019AnES}) that could be interpreted as the day to day patterns of our problem. Once the metadomel is trained using a dataset built using TRNSYS,  all the parameters of a real building and of its Building Management System (BMS) are estimated using real measurements with the Covariance Matrix Adaptation Evolutionary Strategy (CMA-ES) \cite{igel:hansen:roth:2007} which provides a derivative free optimization procedure. A multi-objective methodology to improve energy efficiency
and maintain thermal comfort is then implemented by
acting only on the BMS. The NSGA-II approach is used to
obtain the Pareto optimal parameters. The performance of
this metamodel  are compared at each step with the usual FFN alternatives, LSTM, and GRU  metamodels.

The paper is organized as follows.  Section~\ref{sec:models} provides all the deep learning architectures used in this paper to build a metamodel and describes the data and  variables used in our metamodel. Section~\ref{sec:calib:optim} illustrates the performance of our metamodel in the calibration and optimization process of a real building. The numerical experiments illustrate how this metamodel ensures a significant gain in energy in comparison to the considered alternatives.

\section{Metamodeling}
\label{sec:models}

\subsection{Notations}
Let $(X_k)_{k\geqslant 0}$ be the state of the building i.e. the inside temperatures and the consumptions of the building management system. The index $k$ denotes time and, in the setting of this paper, data are collected each hour. The aim of the metamodel introduced in this paper is to provide a numerically efficient solution to predict $(X_k)_{k\geqslant 0}$ from other variables and external observations such as meteorological data. Such a metamodel is described by several sets of input variables. A parameter $\theta_{\mathrm{build}}$ containing all unknown parameters useful for the geometrical description of the buildings (windows area ratio, etc.) and parameters related to heat transfer  (capacitance, airchange infiltration, etc.). Choosing such parameters allows to build a data set and design a metamodel able to mimick various buildings. A sequence $(W_k,O_k,I_k)_{k\geqslant 0}$ providing at each hour the building management system variables in $I_k$ (comfort and reduced temperatures for the HVAC), the occupancy $O_k$ (described as a percentage of a given maximum number of people) and in $W_k$ the weather data at time $k$. In this section, we describe how a simulation program may be used to train the metamodel which aims at mimicking the outputs of this simulation program for  various choices of $\theta_{\mathrm{build}}$, $(I_k)_{k\geqslant 0}$, $(O_k)_{k\geqslant 0}$ and of meteorological data $(W_k)_{k\geqslant 0}$. The appendix displays a complete list is of the variables contained in $\theta_{\mathrm{build}}$, $(I_k)_{k\geqslant 0}$, $(O_k)_{k\geqslant 0}$ and in $(W_k)_{k\geqslant 0}$ for the numerical experiment of this paper.

\subsection{Models}
In most recent works, a great deal of research activities focused on FFN as surrogate models, \cite{Bre2020AnEM,Magnier2010MultiobjectiveOO, Reynolds2018AZB}. Although they may lead to interesting performance during the training phase, these fully connected architectures are not well suited for time series prediction, in particular for long time spans. We ceased this opportunity to explore other approaches that have proven to be more relevant for solving time series tasks in the past few years. Therefore, we decided to evaluate the go-to architectures for time series: a standard LSTM, a bidirectional GRU (BiGRU), a hybrid model mixing both convolutional and GRU layers (ConvGru), and a Feed Forward Network (FFN) as used in previous works.  In addition to those models, a Transformer model which  introduces an attention mechanism to model dependencies is also considered. These models have been implemented using the deep learning framework PyTorch and can be found on our Github\footnote{https://pytorch.org and https://github.com/maxjcohen/transformer}.

Recurrent Neural Network (RNN) were first introduced as a more suited architecture for dealing with time varying input patterns \cite{Mozer1989AFB}. By replacing buffer based approaches with an updated context state, RNN are able to solve time series problems with short time dependencies, but are lackluster in problems requiring long term memory due to vanishing and exploding gradient \cite{Bengio1994LearningLD}. Long Short Term Memory proposed in \cite{Hochreiter1997LongSM} aim at bridging that gap by enforcing error flow throughout time in the network. Later, the authors of \cite{Cho2014LearningPR} modified the LSTM architecture in order to simplify implementation and improve computation times, resulting in a novel model called Gated Recurrent Unit.
    
In parallel to these advances on recurrent architectures, Convolutional Neural Networks (CNN), rendered popular by \cite{Krizhevsky2012ImageNetCW} for image classification, have been adapted to time series problem. The approaches proposed in \cite{Jzefowicz2016ExploringTL,Kim2016CharacterAwareNL} outperformed traditional Natural Language Processing (NLP) models by replacing the embedding layer with a character-level convolutional layer. Following this idea, \cite{Oord2016WaveNetAG} considerably improved on the speech to text state of the art, by using dilated convolutions, increasing the receptive fields of WaveNet at each layer. One year later, \cite{Oord2017ParallelWF} improved on the existing architecture by introducing Parallel WaveNet, which provided similar performance for a lower computational cost.

Recurrent and convolutional approaches coincide in that temporally close time steps data are matched together. In 2017, \cite{Vaswani2017AttentionIA} proposed an attention based approach to solving NLP tasks that consider the entire input sequence in parallel. The Transformer model is based on a self-attention mechanism, that computes an attention value for every element of a sequence with respect to all others to model their dependency. This attention mechanism allows to understand at each time step $k$ which input elements are crucial to predicting the new state $X_k$. This makes these networks more interpretable than their most widely-used recurrent counterparts such as LSTM or GRU networks and motivate a keen interest for such approach to predict complex time series.

Transformer differ from sequential architectures in that they compute prediction at each time step in parallel. In our context, we propose to use a Transformer architecture as follows. Let $F_{\theta_{\mathrm{meta}}}$ be the Transformer mapping which computes a prediction  $(\widehat X_{k})_{1\leqslant k\leqslant n}$ of the states $(X_{k})_{1\leqslant k\leqslant n}$:
$$
(\widehat X_1,\ldots,\widehat X_n) = F_{\theta_{\mathrm{meta}}}(\theta_{\mathrm{build}}, (W_k,O_k,I_k)_{1\leqslant k\leqslant n})\,,
$$
where $\theta_{\mathrm{meta}}$ contains all the unknown parameters specific to the metamodel. Following the state of the art in sequence to sequence modeling, the Transformer  adopts an encoder-decoder architecture. The encoder computes a latent vector from the input data, which is fed to the decoder in order to predict the outputs. These sub-networks are trained jointly and are supposed to foster learning of a meaningful representation of the data. The encoder and decoder consist of a self attention block, responsible for leveraging the relationship between time steps in the sequence, and a feed forward network, which contains the non linearity of the Transformer.

\paragraph{Embedding.}
Similarly to the original embedding layer,  our metamodel is first based on a linear map, that allows setting the dimension $d_{\mathrm{emb}}$ of the latent representation of the inputs. 
Let $\Delta >0$ be an  attention window and $k$ be a given time step. For all $k-\Delta \leqslant j\leqslant k+\Delta$, let $U_j = (\theta_{\mathrm{build}}, I_j, W_j, O_j)\in\mathbb{R}^d$ be the vector at time $j$ which stacks all inputs, and $U^{\mathrm{emb}}_j$ the latent vector for the corresponding time step,
$$
U^{\mathrm{emb}}_j = W_{\mathrm{emb}} \cdot U_j + b_{\mathrm{emb}}\eqsp,
$$
where $W_{\mathrm{emb}}\in\mathbb{R}^{d_{\mathrm{emb}}\times d}$ and $b_{\mathrm{emb}}\in\mathbb{R}^{d_{\mathrm{emb}}}$ are the unknown weight matrix and bias respectively, that are estimated during the training phase.

\paragraph{Encoder.}
The encoder block proceeds by computing the query, key and value from $R_j$ for this state with a linear transform: 
\begin{equation}
\label{eq:qkv}
q_j =  W^{q} U^{\mathrm{emb}}_j\eqsp,\quad \kappa_j =  W^{\kappa} U^{\mathrm{emb}}_j\eqsp,\quad v_j =  W^{v} U^{\mathrm{emb}}_j \eqsp,
\end{equation}
where $W^{q}$, $W^{\kappa}$ and $W^{v}$ are the unknown $r\times d_{\mathrm{emb}}$ matrices (parameters of the metamodel to be estimated, $r$ chosen by the user). Then, let $K_k$ denote the matrix whose columns are $\kappa_j$, $k-\Delta\leqslant j\leqslant k + \Delta$,  and compute for all $k-\Delta\leqslant j\leqslant k+\Delta$,
$$
s_k^{j} = q_j^TK_k \quad\mbox{and}\quad \pi_k^{j}= \sigma(s_k/\sqrt{r})_j\eqsp,
$$
where $\sigma$ is the softmax function. Finally, self-attention is computed as
\begin{equation}
\label{eq:z}    
z^{\mathsf{enc}}_k =  \sum_{\ell=k-\Delta}^{k+\Delta}\pi_k^{\ell}v_\ell\eqsp.
\end{equation}
The output of the encoder is then given by a final transform of $z^{\mathsf{enc}}_k$ which is considered as the input of a FFN:
$$
r^{\mathsf{lat}}_k = \texttt{FFN}_{\theta_{\mathrm{att}}}(z^{\mathsf{enc}}_k)\eqsp,
$$
where $\theta_{\mathrm{att}} = \{W_1,b_1,W_2,b_2\}$ and 
$$
\texttt{FFN}_{\theta_{\mathrm{att}}}(z) = W_2 \cdot max(0, W_1 \cdot z + b_1)  + b_2\,.
$$
In practice, the self attention computation \eqref{eq:z} is replicated $h$ times, each referred to as "head", that are concatenated before being fed to the FFN. Having multiple heads, i.e. computing multiple instances of self attention in parallel, allows the transformer to set attention to multiple aspect of the input sequence at the same time. In a multi-layer Transformer, the output of each layer is used as an input for the next layer before being processed similarly.

\paragraph{Decoder.} The decoder block acts similarly, except for one added attention step where the keys and values are computed from the latent vectors $r^{\mathsf{lat}}_j$, $k-\Delta\leqslant j \leqslant k+\Delta$.  This produces a vector $z^{\mathsf{enc}}_k$ as in \eqref{eq:z} which is a mixture of the values associated with the latent vectors. This mixture is fed to a FFN to produce $\widehat X_k$.  The parameters to train are therefore $W_{\mathrm{emb}}$, $b_{\mathrm{emb}}$, $\theta_{\mathrm{att}}$ and $W^{q}$, $W^{\kappa}$ and $W^{v}$.

\subsection{Training and validation}
\label{sec:training}
The first step consists in sampling a dataset with TRNSYS to learn the metamodel and  defining ranges for each input parameters in $\theta_{\mathrm{build}}$, $(I_k)_{k\geqslant 0}$ and $(O_k)_{k\geqslant 0}$ with the help of energy managers, such as highest and lowest scheduled temperature, or the most early and late hour of arrival of occupants, see the appendix for a complete list of these ranges. In addition, real weather data $(W_k)_{k\geqslant 0}$ acquired between May and December 2019 where used to obtain a dataset consistent with the real building. As discussed in the previous section, some related papers use Latin Hypercube sampling, introduced in \cite{McKay2000ACO}, to form their dataset. In our numerical experiments, we chose instead a uniform sampling method over the ranges of each variable. This allows us to easily split the dataset into $k$-folds, which will be useful for the validation step discussed in the next section.

During this step, daily values defined in the appendix are converted to a time series whose value changes with every day. This way, there are 38 variables in the input vector at each time step: 19 variables from $\theta_{\mathrm{build}}$, 7 from $W_k$, 1 from $O_k$ and 11 from $I_k$. A total of 38000 training examples were sampled, an example being a week i.e. 168 hours. 
During the training phase, the parameters of each metamodel described in Section~\ref{sec:models} are estimated based on this dataset (called  $\theta_{\mathrm{meta}}$ in the detailed case of the Transformer approach). The metamodels compared in this section are defined with a latent dimension of $d_{emb} = 64$ and a total of $N=8$ layers. These values were obtained through a grid search, see the appendix for additional information. Other hyper parameters, such as learning rate dropout, number of epochs or batch size, were chosen empirically.
    
During training, for each example, we use a loss function defined by Energy Management experts, consisting of a combination between mean squared consumption and temperature errors:
\begin{align*}
\Delta_T^{\theta_{\mathrm{meta}}} = \left(\frac{1}{N}\sum_{k=1}^N (\widehat T_k^{\theta_{\mathrm{meta}}} - T_k)^2\right)^{1/2}\quad\mathrm{and}&\quad
\Delta_Q^{\theta_{\mathrm{meta}}} = \left(\frac{1}{N}\sum_{k=1}^N (\widehat Q_k^{\theta_{\mathrm{meta}}} - Q_k)^2\right)^{1/2}\,, \\
\mathrm{loss}(\theta_{\mathrm{meta}}) = \alpha\log(1 + \Delta_T^{\theta_{\mathrm{meta}}}) &+ \beta \cdot \log(1 + \Delta_Q^{\theta_{\mathrm{meta}}})\,,
\end{align*}
where $N$ is the number of data in each example, $T_k$ and $Q_k$ are the ground truth at time $k$, and $\widehat T_k^{\theta_{\mathrm{meta}}}$ and $\widehat Q_k^{\theta_{\mathrm{meta}}}$ are the predictions given by the metamodel with the current value $\theta_{\mathrm{meta}}$ of the metamodel for temperature and consumption respectively. In this experiments below, we chose $\alpha=1$ and $\beta=0.3$. We chose the Adam optimizer \cite{Kingma2015AdamAM} ; all simulations were computed on a single 1080TI GPU card. Table~\ref{table:train} displays the mean values and standard deviations of the loss function on the validation dataset after training. The table also displays the mean squared error $\mathrm{MSE}_T$ (resp. $\mathrm{MSE}_Q$) on the temperatures (resp. consumptions) only, and these metrics computed only during occupation time $\mathrm{MSE_T^{occ}}$ and $\mathrm{MSE_Q^{occ}}$.  In addition, the coefficients of determination (rescaled  mean squared errors relative to the
target data) of the temperatures $R^2_T$ and consumptions $R^2_Q$ are given. These coefficients of determination are computed with the Python function {\em sklearn.metrics.r2\_score}.

\begin{table}
\caption{Metrics (means and standard deviations) of the metamodels on the validation dataset. The best mean values are displayed in bold (the lowest losses and mean squared errors and the coefficient of determination closest to 1).}
\label{table:train}
\centering 
\begin{tabular}{*7c}\toprule
& Transformer & BiGRU & LSTM & ConvGru & FFN \\
\toprule
$\mathrm{Loss}$ \;\;\;\;\;($\times10^{-4}$)& $\textbf{1.13\eqsp (0.746)}$ & $1.43\eqsp (1.06)$ & $13.8\eqsp (4.55)$ & $2.78\eqsp (1.77)$ & $61.1 (27.4)$ \\
%\hline
$\mathrm{MSE_T}$ \;\;($\times10^{-5}$) & $\textbf{3.86\eqsp (4.53)}$ & $4.28\eqsp (5.18)$ & $7.32\eqsp (7.75)$ & $9.37\eqsp (11.5)$ & $178 (205)$ \\
%\hline
$\mathrm{MSE_Q}$ \;\;($\times10^{-4}$)& $\textbf{2.47\eqsp (2.30)}$ & $3.34\eqsp (2.98)$ & $43.7\eqsp (14.7)$ & $6.16\eqsp (4.30)$ & $146 (54.2)$ \\
%\hline
$\mathrm{MSE_T^{occ}}$ ($\times10^{-5}$) & $\textbf{1.08\eqsp (1.32)}$ & $1.18\eqsp (1.54)$ & $2.02\eqsp (2.52)$ & $2.77\eqsp (3.37)$ & $51.2 (64.1)$ \\
%\hline
$\mathrm{MSE_Q^{occ}}$  ($\times10^{-4}$)& $\textbf{1.06\eqsp (1.29)}$ & $1.21\eqsp (1.92)$ & $3.61\eqsp (2.93)$ & $2.28\eqsp (2.35)$ & $43.2 (25.1)$ \\
%\hline
\emph{$R^2_T$} \;\;\;\;\;\;\;($\times10^{-3}$)& $\textbf{996\eqsp (0.832)}$ & $996\eqsp (1.40)$ & $992\eqsp (1.64)$ & $990\eqsp (2.10)$ & $829 (43.2)$ \\
%\hline
\emph{$R^2_Q$} \;\;\;\;\;\;\;($\times10^{-3}$)& $\textbf{760\eqsp (240)}$ & $657\eqsp (593)$ & $559\eqsp (473)$ & $707 (268)$ & $-738 (3080)$ \\
\bottomrule
\end{tabular}
\end{table}
    
\section{Energy Optimization in a real building}
\label{sec:calib:optim}
The experiments conducted in our paper to analyze the performance of the metamodel trained in Section~\ref{sec:training} focused on the optimization of a $28733\, m^2$ building located in the Parisian region. The total building is represented by a single thermal zone including  5 vertical walls with respective following areas $3521\, m^2$, $2692\, m^2$, $3257\, m^2$, $599\, m^2$ and $16329\, m^2$, a horizontal roof and a horizontal ground. Based on a commonly used rule, it is assumed that $2/3$ of the full area is occupied by people. Assuming that each occupant requires $12\, m^2$, this allows to set the initial values for the number of occupants and the number of PCs (set to 1.2 times this value) in the building during occupancy hours. These values are assumed to be known and fixed and used to sample the training dataset.

\subsection{Calibration}
During the training phase, metamodel parameters are estimated by minimizing the loss function on the simulated dataset which corresponds to various choices of $\theta_{\mathrm{build}}$, $(I_k,O_k,W_k)_{k\geqslant 0}$, associated with building behaviors $(X_k)_{k\geqslant 0}$. This metamodel has been trained on a dataset containing only simulated data,  ignoring real building related noise and measurement errors. Additionally, both the BEM and our surrogate model take as input a number of variables, such as window to wall ratio, window leakage, or wall construction, that cannot be properly identified for each building. By comparing the metamodel predictions to real historic data during the calibration phase, we search for a set of building related parameters that best match reality.
    
During this step, the weights $\theta_{\mathrm{meta}}$ of the metamodel are frozen, meaning that we no longer back propagate the error, nor do we update each weight matrix of the neural network. Using the coefficient of determination as a cost function, we can compute, for each given set of input parameters $\theta_{\mathrm{build}}$, $(I_k,O_k,W_k)_{k\geqslant 0}$, the difference between estimated and real historical data. Because this is a non differentiable problem, the cost function cannot be minimized using the same algorithm as in the training step; instead we use the CMA evolution strategy (CMA-ES, \cite{Hansen2016TheCE}), an evolutionary algorithm adapted to derivative free non-convex optimisation problems in continuous domain. It is implemented by the author of the paper in the pycma library\footnote{https://github.com/CMA-ES/pycma}.

\begin{figure}
\centering
\includegraphics[width=0.95\textwidth]{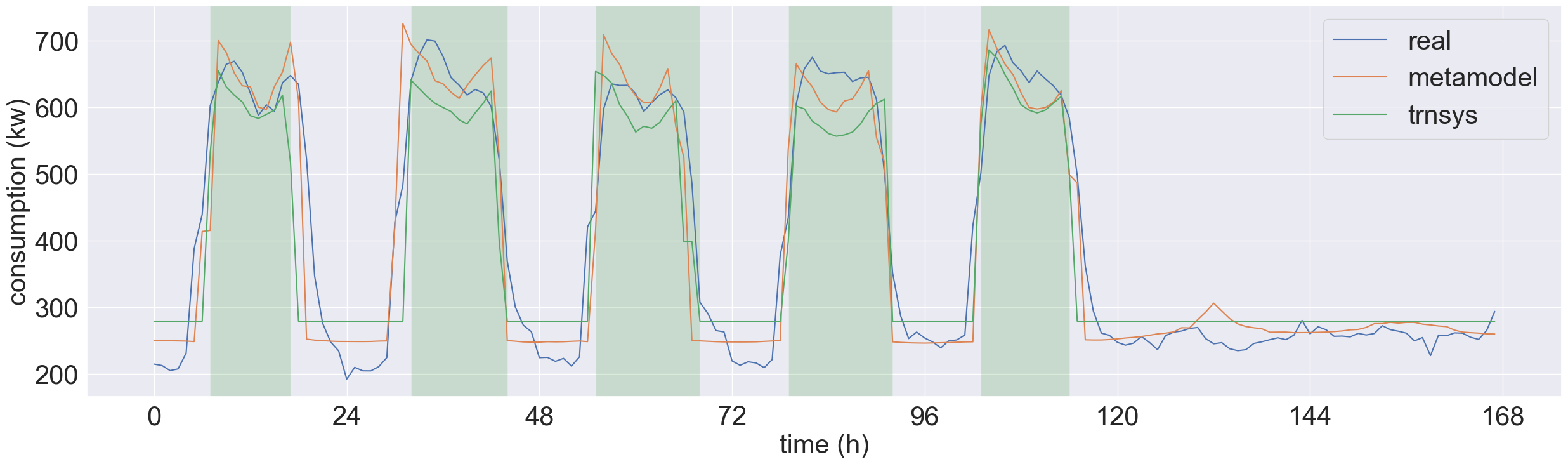}
\includegraphics[width=0.95\textwidth]{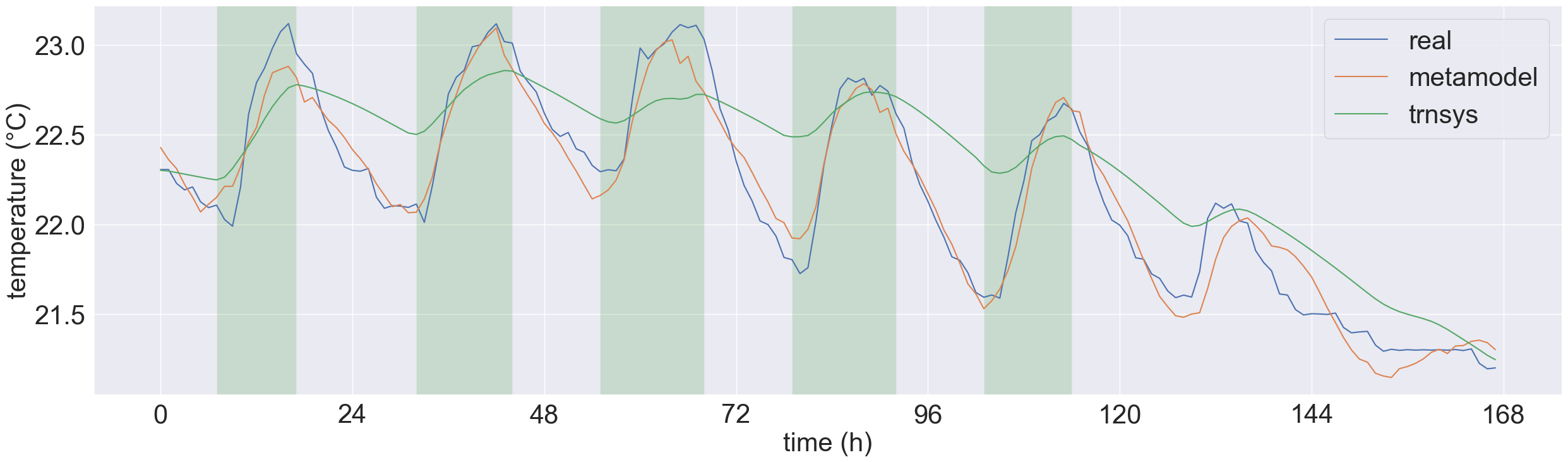}
\includegraphics[width=0.95\textwidth]{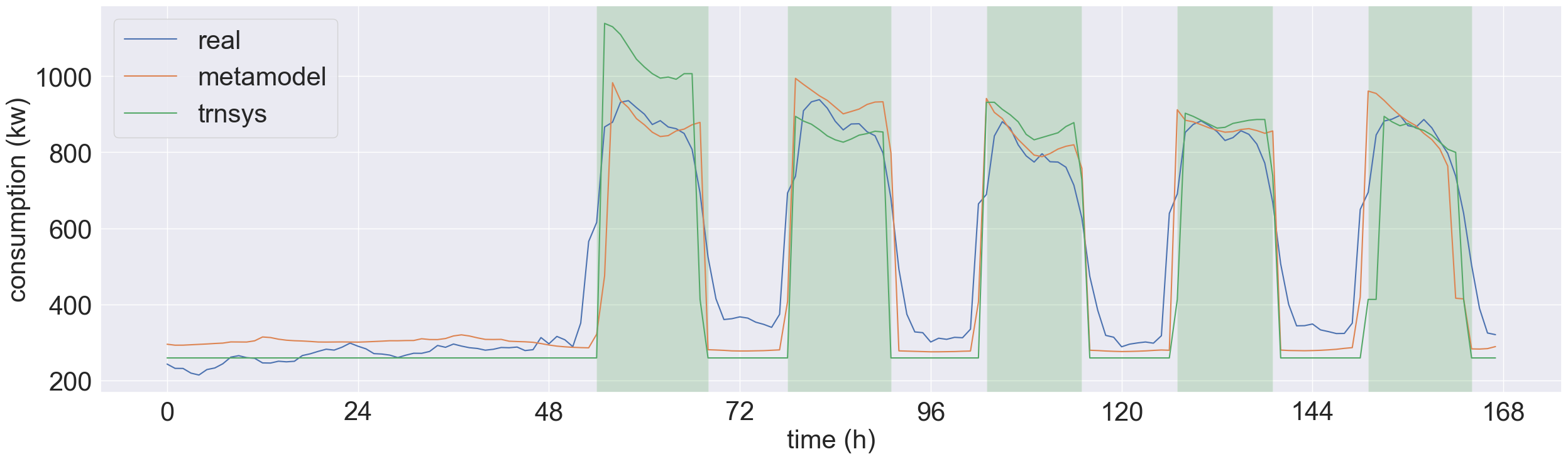}
\includegraphics[width=0.95\textwidth]{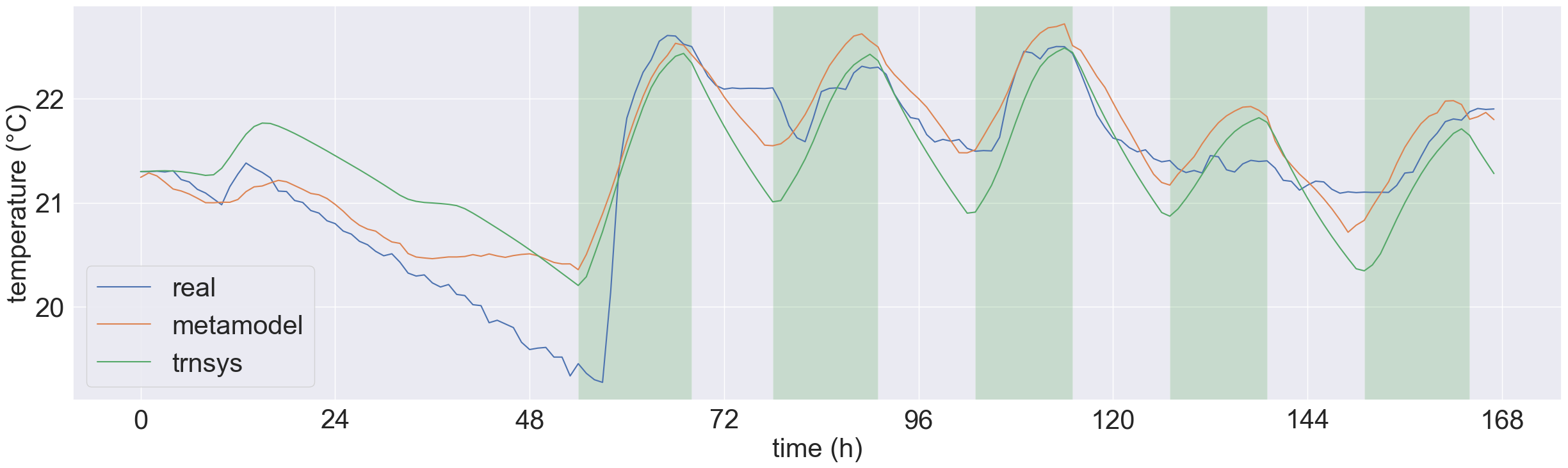}
\caption{Consumption and temperature simulations after calibration, for both the metamodel and TRNSYS, for week 1 (top) and week 2 (bottom). Green bars indicate occupation periods.}
\label{fig:calib}
\end{figure}

Calibration was run until convergence for the metamodel, and for a maximum of 8 hours for the original BEM (TRNSYS). We can see the advantage of going through the training of a metamodel when comparing a calibration for both TRNSYS and the metamodel, as we are now able to reach lower costs in a much shorter time frame. This is confirmed by Table~\ref{tab:calib} which displays the Mean squared error for the temperatures and heating consumption after calibration using TRNSYS and the Transformer-based metamodel, for two different weeks shown in Figure~\ref{fig:calib}.

\begin{table}
    \caption{Metrics after calibration for two weeks, beginning the 4th and the 30th of November 2019. Calibration run for 500 epochs (resp. 2500 epochs) for the metamodel (resp. for TRNSYS).}
    \label{tab:calib}
    \centering 
    \begin{tabular}{*9c}    \toprule
    & $\mathrm{MSE_T}$ & $\mathrm{MSE_Q}$ & $\mathrm{MSE_T^{occ}}$ & $\mathrm{MSE_Q^{occ}}$ & \emph{$R^2_T$} & \emph{$R^2_Q$} & time (h) \\\toprule
    {\bf Week 1} & & & & & & & & \\
    TRNSYS & $1.04\cdot10^{-1}$ & $4967$ & $3.31\cdot10^{-2}$ & $1434$ & 0.644 & 0.848 & 2 \\
    Metamodel & $1.62\cdot10^{-2}$ & $3241$ & $4.71\cdot10^{-3}$ & $477$ & 0.945 & 0.901 & 2 \\
    \midrule
    {\bf Week 2} & & & & & & & & \\
    TRNSYS & $2.66\cdot10^{-1}$ & $16067$ & $6.58\cdot10^{-2}$ & $6782$ & 0.592 & 0.761 & 2 \\
    Metamodel & $1.42\cdot10^{-1}$ & $10493$ & $6.55\cdot10^{-2}$ & $5162$ & 0.782 & 0.844 & 2 \\
    \bottomrule
    \end{tabular}
\end{table}

\subsection{Optimization}
\begin{figure}
\centering
\includegraphics[width=0.5\textwidth]{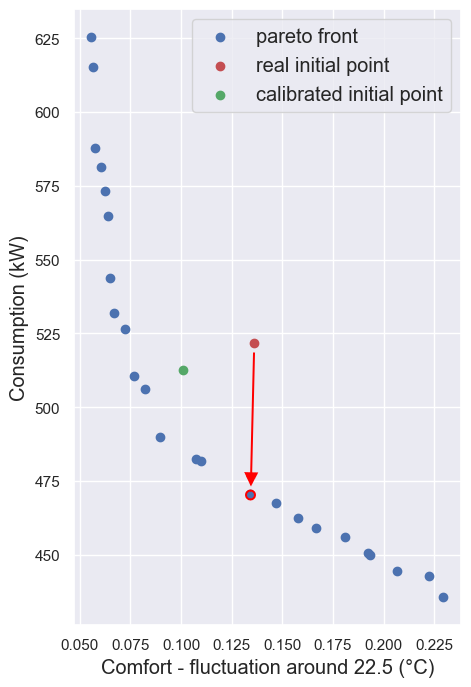}
\caption{Pareto front after optimization for the second week. We select the point of closest equivalent comfort, corresponding to a 9.31\% reduction in consumption.}
\label{fig:pareto}
\end{figure}

\begin{figure}
\centering
\includegraphics[width=0.95\textwidth]{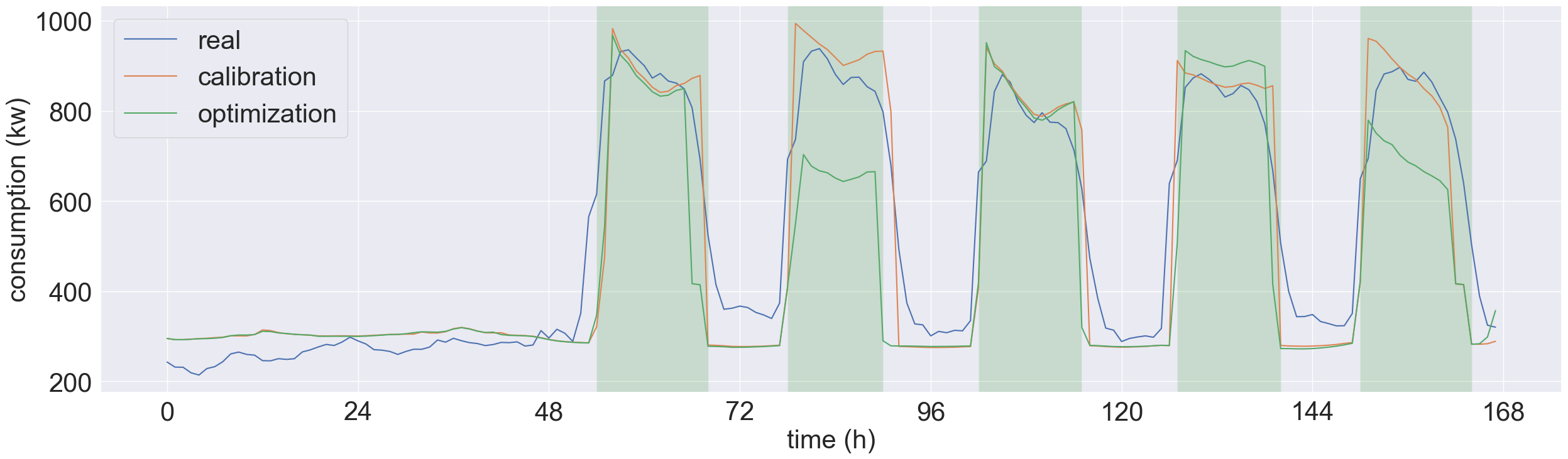}
\includegraphics[width=0.95\textwidth]{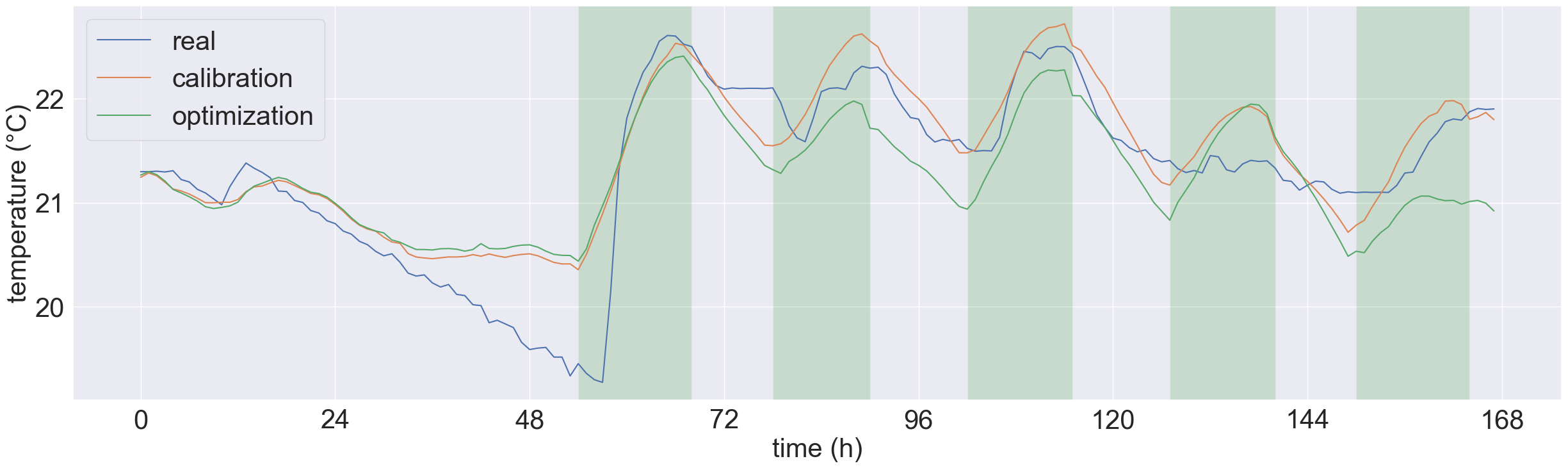}
\caption{Consumption and temperature simulations after optimization (metamodel) for the second week. Green bars indicate occupation periods.} 
\label{fig:timeseriesafteroptim}
\end{figure}

Once the metamodel is calibrated, we can use it as an accurate simulator for how the building will react to changes in its usage. After a successful calibration, all building related variables contained in $\theta_{\mathrm{build}}$ are correctly estimated. The parameters $I_k$ associated with the HVAC system can be optimized for a given set of weather data $W_k$. The optimization tasks consists in finding a set a usage related parameters that reduce consumption while keeping the same level of comfort. Optimizing energy consumption requires minimizing two conflicting objectives, making it impossible to find a solution that optimize both objectives simultaneously. Instead, we search for optimal compromises between energy consumption and comfort, in the form of a Pareto front. Indeed, for any such optimal compromise, we can always get a higher level of comfort, for the price of a higher consumption. The consumption criteria is the energy load during the week ; the comfort criteria is the gap between indoor temperature and a constant reference temperature $T^*$:
\begin{align*}
    \Delta^{\mathrm{opt}}_T = \frac{1}{N^{\mathrm{opt}}_{Occ}}\left(\sum_{k=1}^{N^{\mathrm{opt}}} \mathds{1}_{k\in \mathrm{Occ}}(\widehat T_k - T^*)^2\right)^{1/2}\quad\mathrm{and}&\quad
    \Delta^{\mathrm{opt}}_Q = \frac{1}{N^{\mathrm{opt}}} \sum_{k=1}^{N^{\mathrm{opt}}} \widehat Q_k\,,
\end{align*}
where $T^*=22.5^{\circ}C$, $N^{\mathrm{opt}}$ is the number of hours to be considered in the optimization process and $\mathrm{Occ}$ is a subset of daytime hours specifying at which hours the target temperature has to be reached in the building. Following recent works in building energy optimization, we search for a set of optimal parameters using NSGA-II (\cite{Deb2000AFE}), another evolutionary algorithm, but adapted to multi objective problems. An implementation can be found in the Pygmo\footnote{https://esa.github.io/pygmo2/} library. In the absence of a stopping condition, we simply run the optimization for a set 3000 epochs (2 hours). The result can be viewed as a Pareto front which is given in Figure \ref{fig:pareto} for the second week used in the calibration process. As observed during calibration, this process can take a colossal number of epochs before achieving satisfactory results, once again justifying the use of a much faster metamodel. The time series of consumption and temperatures associated with the BMS parameters selected in Figure~\ref{fig:pareto} are given in Figure~\ref{fig:timeseriesafteroptim}.

\section{Conclusion}
In this paper, we proposed an end-to-end metamodeling methodology to optimize building energy loads and to reduce computational costs. The proposed metamodel ensures compatibility between simulations and real building observations through a calibration step. We experimented with various deep learning architectures more suited to recurrent problems than Feed Forward Networks. Results show that a wide variety of models display encouraging results on our sampled dataset, while largely outperforming FFN. During optimization, we chose to maintain the same level of comfort as the historical data, in order to have as little impact as possible on the working environment. Compared to calibrated simulations, we were able to reduce consumptions significantly.

\appendix

\section{Dataset}
\subsection{Inputs}
    The dataset is divided in four sub-variables, each representing a different aspect of the simulation. When creating the dataset using the Building Energy Model (here TRNSYS), we sampled each variable uniformly in a given interval. These intervals  are also used for the calibration process.
    
    \begin{enumerate}[-]
        \item $\theta_{building}$ represents the geometric properties of the building, see Table~\ref{tab:theta-building}.
        \item $I_k$ stores the schedules and settings of the heating and ventilation, see  Table~\ref{tab:Ik}. This is the only parameter tuned during the optimization process.
        \item $O_k$ stores the occupation schedules of the building, see Table~\ref{tab:Ok}. These values are calibrated to match real data, and kept fixed during the optimization.
        \item $W_k$ stores the weather data for the week, see Table~\ref{tab:Wk}. In this paper, we exclusively use data collected by weather institutes, that are available afterward. In practice, these data would be replaced by weather forecasts.
    \end{enumerate}
    
\begin{figure}
        \centering
        \includegraphics[width=\textwidth]{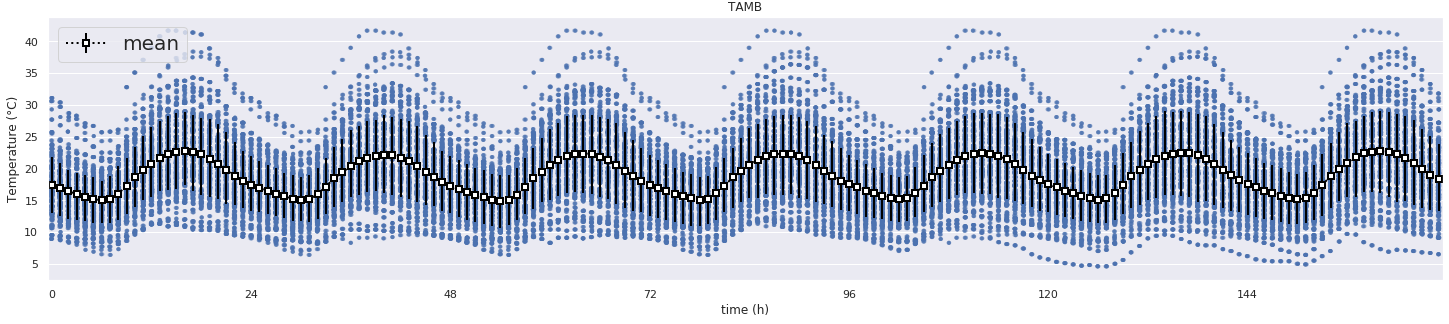}
        \caption{Distribution of the outdoor temperature in the dataset, stored in the \texttt{T\_AMB} variable in the Table~\ref{tab:Wk}. Squares indicates the mean value, while vertical bars represent 85\% of the data.}
        \label{fig:dataset_input_distribution}
    \end{figure}

    The distribution of the outdoor temperature can be found in Figure~\ref{fig:dataset_input_distribution}. We sampled a total of 40 000 examples for our dataset, of which 38 000 were used for training, 1 000 for validation and 1000 for testing purposes.
    
\begin{table}
    \centering
    \begin{tabular}{@{}llll@{}}
    Variable & Minimum & Maximum & Step \\ \midrule
	airchange\_infiltration\_vol\_per\_h & 0.1 & 0.5 & 0.1 \\ 
	capacitance\_kJ\_perdegreK\_perm3 & 50 & 300 & 10 \\ 
	power\_VCV\_kW\_heat & 0 & 1000 & 100 \\ 
	power\_VCV\_kW\_clim & 0 & 1000 & 100 \\ 
	nb\_occupants & 1000 & 2000 & 200 \\ 
	nb\_PCs & 1000 & 2000 & 200 \\ 
	percent\_light\_night & 0 & 70 & 10 \\ 
	percent\_PCs\_night & 0 & 70 & 10 \\ 
	facade\_1\_thickness\_2 & 0.05 & 0.15 & 0.05 \\ 
	facade\_2\_thickness\_2 & 0.05 & 0.15 & 0.05 \\ 
	facade\_3\_thickness\_2 & 0.05 & 0.15 & 0.05 \\ 
	facade\_4\_thickness\_2 & 0.05 & 0.15 & 0.05 \\ 
	roof\_1\_thickness\_3 & 0.05 & 0.15 & 0.05 \\ 
	facade\_1\_window\_area\_percent & 40 & 50 & 5 \\ 
	facade\_2\_window\_area\_percent & 40 & 50 & 5 \\ 
	facade\_3\_window\_area\_percent & 40 & 50 & 5 \\ 
	facade\_4\_window\_area\_percent & 40 & 50 & 5 \\ 
    \bottomrule \\
    \end{tabular}
    \caption{$\theta\_buidling$ ranges.}
    \label{tab:theta-building}

\end{table}

\begin{table}
    \centering
    \begin{tabular}{@{}llll@{}}
    Variable & Minimum & Maximum & Step \\ \midrule
	start\_clim\_day & 7 & 9 & 1 \\
	end\_clim\_day & 18 & 20 & 1 \\ 
	t\_clim\_red\_day & 24 & 30 & 0.5 \\ 
	t\_clim\_conf\_day & 20 & 24 & 0.5 \\ 
	start\_heat\_day & 6 & 8 & 1 \\ 
	end\_heat\_day & 17 & 19 & 1 \\
	t\_heat\_red\_day & 17 & 22 & 0.5 \\
	t\_heat\_conf\_day & 22 & 24 & 0.5 \\ 
	start\_ventilation\_day & 7 & 9 & 1 \\ 
	end\_ventilation\_day & 18 & 20 & 1 \\ 
	t\_ventilation\_day & 18 & 26 & 0.5 \\ 
	vol\_ventilation\_day & 0.7 & 1.7 & 0.3 \\ 
    \bottomrule \\
    \end{tabular}
    \caption{$I_k$ ranges. Each parameter can hold a different value for each day of the week. For ease of reading, we replaced them by a single line, as the ranges are the same for every day.}
    \label{tab:Ik}

\end{table}

\begin{table}
    \centering
    \begin{tabular}{@{}llll@{}}
    Variable & Minimum & Maximum & Step \\ \midrule
	start\_occupation\_monday & 7 & 9 & 1 \\ 
	start\_occupation\_tuesday & 7 & 9 & 1 \\ 
	start\_occupation\_wednesday & 7 & 9 & 1 \\ 
	start\_occupation\_thursday & 7 & 9 & 1 \\ 
	start\_occupation\_friday & 7 & 9 & 1 \\
	end\_occupation\_monday & 17 & 20 & 1 \\ 
	end\_occupation\_tuesday & 17 & 20 & 1 \\ 
	end\_occupation\_wednesday & 17 & 20 & 1 \\ 
	end\_occupation\_thursday & 17 & 20 & 1 \\ 
	end\_occupation\_friday & 17 & 20 & 1 \\
    \bottomrule \\
    \end{tabular}
    \caption{$O_k$ ranges.}
    \label{tab:Ok}
\end{table}

\begin{table}
    \centering
    \begin{tabular}{*2c}
    Variable & Description \\ \midrule
	DNI & Direct Normal Irradiance\\ 
	IBEAM\_H & Direct Horizontal Irradiance \\ 
	IBEAM\_N & Direct Normal Irradiance \\ 
	IDIFF\_H & Diffuse Horizontal Irradiation \\ 
	IGLOB\_H & Global Horizontal Irradiance \\ 
	RHUM & Humidity \\ 
	TAMB & Outdoor temperature \\ 
    \bottomrule \\
    \end{tabular}
    \caption{Weather data as contained in $W_k$.}
    \label{tab:Wk}
\end{table}
    
\subsection{Outputs}
\begin{figure}
        \centering
        \includegraphics[width=0.75\textwidth]{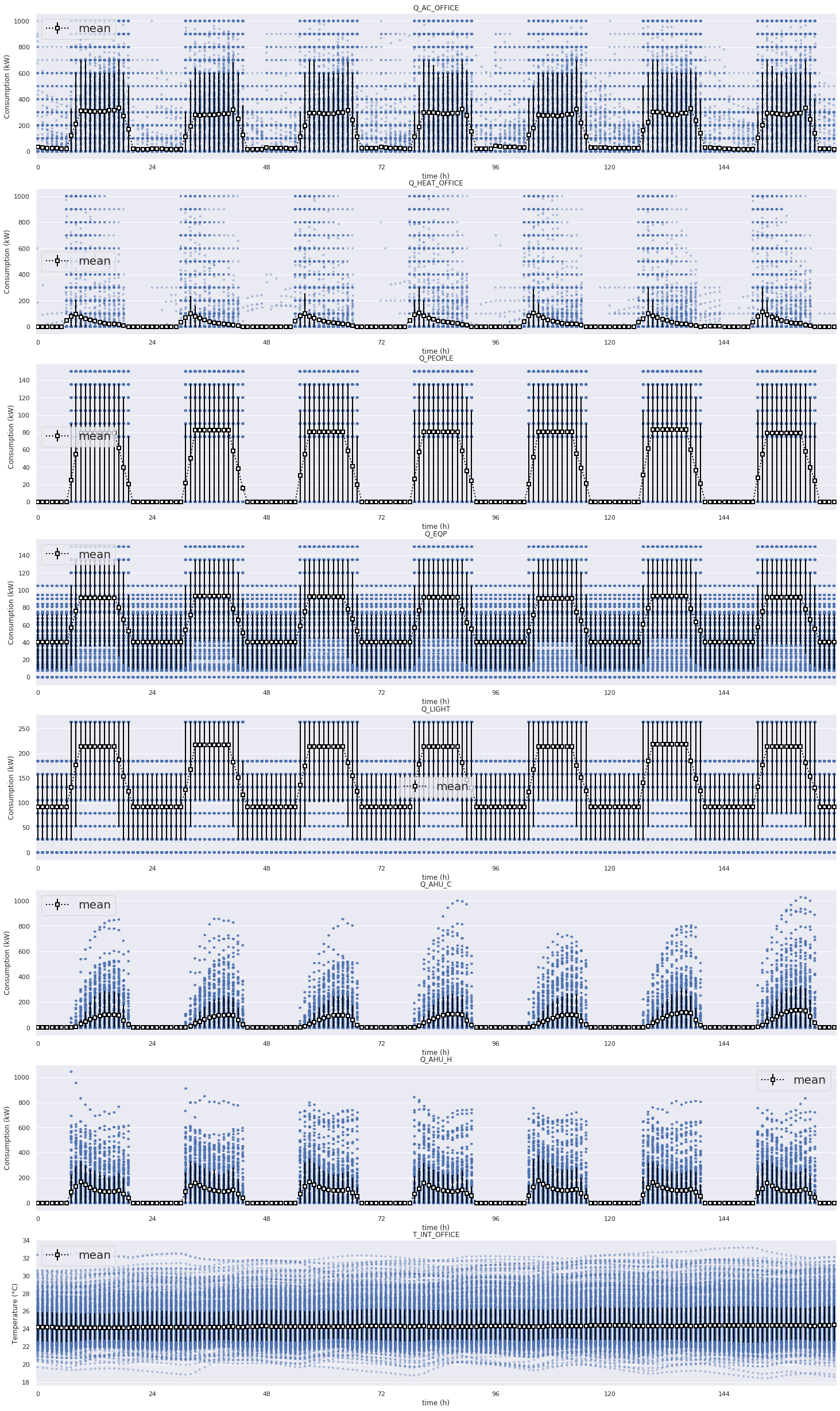}
        \caption{Distribution of the output variables of the BEM. See Table~\ref{tab:output_variables} for a exhaustive list and description of each one. Squares indicates the mean value, while vertical bars represent 85\% of the data.}
        \label{fig:dataset_output_distribution}
    \end{figure}

    The BEM outputs 8 simulated variables at each time step, representing inside temperature as well as various consumption. These variables are aggregated differently during calibration and optimization. Since the metamodel aims at replicating the BEM behavior, it is trained to output these same 8 variables. See Table~\ref{tab:output_variables} for a exhaustive list and description of each one. Their distributions in the dataset can be found in Figure~\ref{fig:dataset_output_distribution}.
    
\section{Metamodel training}
    In order to find an optimal set of hyper parameters, we conducted a grid search for the Transformer model. A list of each parameter, along with its ranges and final value, can be found in Table~\ref{tab:grid_search}.
    
    \begin{table}
        \centering
        \begin{tabular}{*3c}
        Variable & Tested values & Chosen value \\ \midrule
    	Latent dimension ($d_{emb}$) & 16, 32, 64, 128 & 64\\ 
    	Queries ($W^q$) and Keys ($W^k$) matrix size & 4, 8, 16 & 8 \\ 
    	Values ($W^v$) matrix size & 4, 8, 16 & 8 \\
    	Number of heads ($h$) & 4, 8, 16 & 8 \\
    	Number of layers ($N$) & 4, 8, 16 & 4 \\
    	Attention size ($\Delta$) & 6, 12, 24 & 12 \\ 
        \bottomrule \\
        \end{tabular}
        \caption{Hyper parameters tuned during grid search, along with their tested and chosen values.}
        \label{tab:grid_search}
    \end{table}

\section{Calibration}
Sensors installed in the building yield two time series.
    \begin{itemize}
        \item \textbf{Indoor temperature}: we average the values from a set of sensors, in order to obtain a unique indoor temperature value at each time step. This temperature is compared to the simulated indoor temperature (\texttt{T\_INT}).
        \item \textbf{Heat consumption}: we define building heat consumption as the sum of multiple private heat consumptions obtained from sensors. This variable contains the heating consumption (corresponding to \texttt{Q\_HEAT\_OFFICE} for the metamodel), as well as the heating AHU consumption (\texttt{Q\_AHU\_HEAT}), and the equipment and lighting consumption (\texttt{Q\_EQP} and \texttt{Q\_LIGHT} respectively), see Table~\ref{tab:output_variables} for a description of the metamodel output variables. These four simulated variables are summed and compared to the real heat consumption.
    \end{itemize}

\begin{table}
    \centering
    \begin{tabular}{@{}llll@{}}
    variable & description\\ \midrule
	Q\_AC\_OFFICE & AC consumption \\ 
	Q\_HEAT\_OFFICE & Heat consumption \\ 
	Q\_PEOPLE & Heating power due to human activities in the building \\ 
	Q\_EQP & Consumption of equipment, such as computers, elevators, fridges \\ 
	Q\_LIGHT & Consumption of lights \\ 
	Q\_AHU\_C & Consumption of AHU when cooling outside air \\ 
	Q\_AHU\_H & Consumption of AHU when heating outside air \\ 
	T\_INT\_OFFICE & Indoor temperature\\ 
    \bottomrule \\
    \end{tabular}
    \caption{BEM's output variables at each time step.}
    \label{tab:output_variables}
    
\end{table}

\clearpage

\bibliographystyle{apalike}
\bibliography{biblio}
\end{document}